\documentclass{INTERSPEECH2023}

\usepackage{epsfig}
\usepackage[ruled,vlined]{algorithm2e}
\usepackage{multirow}
\usepackage{makecell}
\usepackage{url}
\usepackage{colortbl}
\usepackage{caption}
\usepackage{subcaption}
\usepackage[table]{xcolor}

\interspeechcameraready

\title{CASEIN: Cascading Explicit and Implicit Control for \\ Fine-grained Emotion Intensity Regulation}

\name{Yuhao Cui$^1$, Xiongwei Wang$^1$, Zhongzhou Zhao$^1$, Wei Zhou$^1$, Haiqing Chen$^1$}
\address{
  $^1$Alibaba Group, China}
\email{cyhbrilliant@gmail.com, xiongwei.wangxw@gmail.com, zhongzhou.zhaozz@alibaba-inc.com, zhouwei546138922@126.com, haiqing.chenhq@alibaba-inc.com}

\begin{document}

\maketitle
 
\begin{abstract}
  Existing fine-grained intensity regulation methods rely on explicit control through predicted emotion probabilities. However, these high-level semantic probabilities are often inaccurate and unsmooth at the phoneme level, leading to bias in learning. Especially when we attempt to mix multiple emotion intensities for specific phonemes, resulting in markedly reduced controllability and naturalness of the synthesis. To address this issue, we propose the CAScaded Explicit and Implicit coNtrol framework (CASEIN), which leverages accurate disentanglement of emotion manifolds from the reference speech to learn the implicit representation at a lower semantic level. This representation bridges the semantical gap between explicit probabilities and the synthesis model, reducing bias in learning. In experiments, our CASEIN surpasses existing methods in both controllability and naturalness. Notably, we are the first to achieve fine-grained control over the mixed intensity of multiple emotions.
\end{abstract}
\noindent\textbf{Index Terms}: Emotional TTS, fine-grained control, emotion intensity control, mixed emotions, manifold learning

\section{Introduction}
\label{sec:introduction}

In controllable expressive end-to-end speech synthesis, two paradigms for expression control are utilized: \textit{explicit control} and \textit{implicit control}. Explicit control methods \cite{younggun_lee_2017_emotional,xiaolian_zhu_controlling,kun_zhou_2022_mixedemo,rui_liu_2021_strengthnet,yiwei_guo_2022_emodiff,yi_lei_2021_finegrained,yi_lei_2022_msemotts} rely on a direct representation, such as emotion class and emotion probability, that the user can manipulate. This high-level semantic representation is ideal for emotion control tasks since it accurately reflects the semantics of the control signal. Implicit control methods \cite{yuxuan_wang_2017_uncovering,yuxuan_wang_2018_gst,younggun_lee_2019_robust,viacheslav_klimkov_2019_finegrained}, however, are based on an implicit representation derived from reference speech, such as style-related and prosody-related vectors. This low-level semantic representation retains more raw information, making it suitable for style transfer tasks.

Our study focuses on fine-grained emotion intensity regulation, which involves continuously adjusting the intensity of a specific emotion expression from 0\% to 100\%, with control over each individual phoneme. We also explore a fine-grained emotion mixing task, which enhances expressiveness by allowing the simultaneous regulation of multiple emotions on each phoneme, beyond just the primary emotion.


Zhu \textit{et al.} \cite{xiaolian_zhu_controlling} used Relative Attributes \cite{devi_parikh_2011_relattr} to learn a rank function, which extracts primary emotion probability as the intensity. Other techniques \cite{rui_liu_2021_strengthnet,yiwei_guo_2022_emodiff} use deep neural networks to predict emotion probability for accurate emotion synthesis. The concept of mixed emotions is introduced in psychology \cite{plutchik_1980_theories,georgios_n_yannakakis_2017_theordinal}, represented as a combination of other emotions, e.g., Proud=90\%Happiness+45\%Surprise. Zhou \textit{et al.} \cite{kun_zhou_2022_mixedemo} built upon this concept by formulating a mixed emotion intensity regulation task, predicting intensity distribution over multiple emotions simultaneously instead of single emotion intensity.



\begin{figure}
  \begin{center}
  \includegraphics[width=0.47\textwidth]{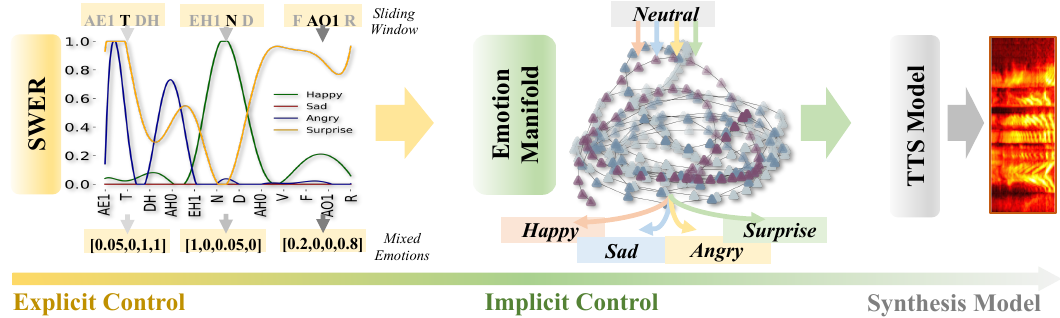}
  \vspace{-18pt}
  \caption{The overview of cascaded control paradigms.}
  \vspace{-33pt}
  \label{fig:intro}
  \end{center}
\end{figure}

\begin{figure*}
  \begin{center}
    \includegraphics[width=1\textwidth]{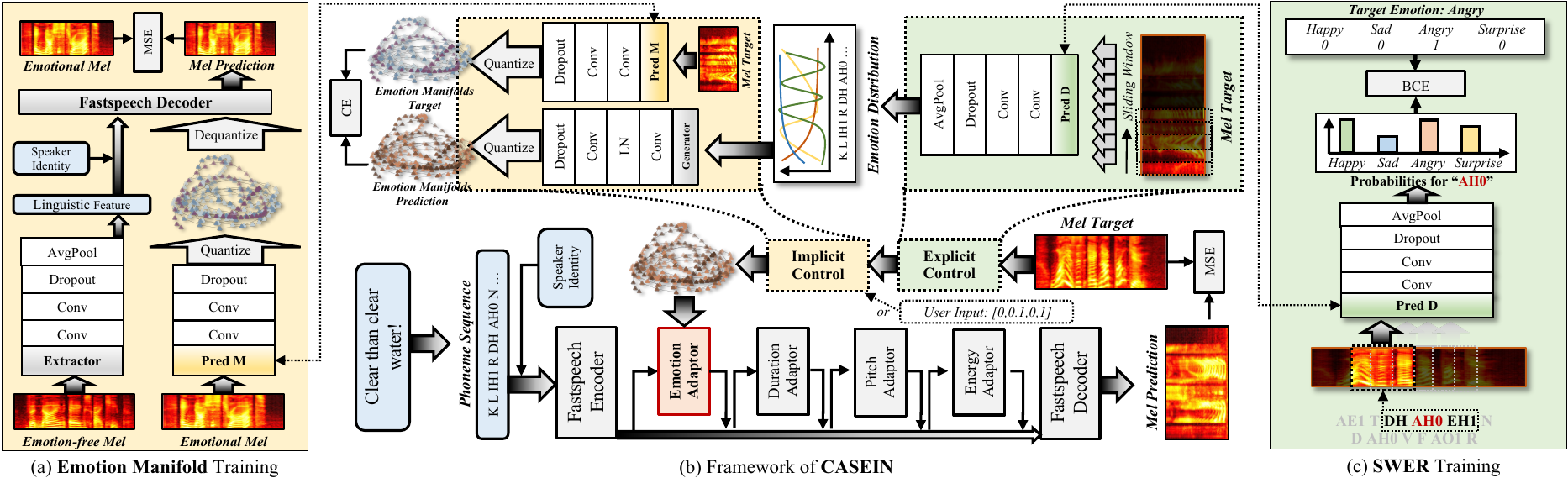}
    \vspace{-17pt}
    \caption{The pipeline for CASEIN training. The Emotion Manifold Predictor (Pred M) and Emotion Distribution Predictor (Pred D) are trained separately and then cascaded during synthesis. The BCE, CE and MSE are loss functions.}
    \vspace{-29pt}
    \label{fig:framework}
  \end{center}
\end{figure*}

However, such coarse-grained emotion control techniques suffer from the limitation of inconsistent and uncontrollable preceding and following segments, especially in mixed emotion synthesis, where coarse-grained control can result in a patchwork of emotions rather than a genuine blend of them. Recent works \cite{yi_lei_2021_finegrained,yi_lei_2022_msemotts} refine the control granularity by using Relative Attributes to extract emotion intensity on each phoneme separately, achieving fine-grained control over the phoneme sequence and enabling more precise emotion synthesis. But our experiments showed that fine-grained control resulted in less natural speech synthesis compared to coarse-grained control. Additionally, regulating mixed emotions using fine-grained control was especially challenging due to the short duration of phonemes, resulting in inaccurate emotion intensity prediction. This difficulty in accurately capturing emotions made it challenging for the synthesis model to bridge the semantic gap and align with high-level semantic representations using explicit fine-grained control.



Fine-grained implicit control involves low-level semantic information, which reduces the semantic gap between the representation and synthesis model. However, prior researches \cite{younggun_lee_2019_robust,viacheslav_klimkov_2019_finegrained,xiang_li_2021_towards} have primarily focused on prosody or style representations that lack direct emotion control by users. While Um \textit{et al.} \cite{seyun_um_2020_emotional} demonstrated the conversion of emotion intensity to a prosody-related vector through clustering interpolation calculations, the non-linear relationship between emotion and prosody often results in imprecise control in implicit schemes.

We propose the \textbf{\underline{CAS}}caded \textbf{\underline{E}}xplicit and \textbf{\underline{I}}mplicit co\textbf{\underline{N}}trol framework, abbreviated as \textbf{CASEIN}, to enhance fine-grained emotion controllability as shown in Figure \ref{fig:intro}. Our approach introduces an implicit \textbf{Emotion Manifold} representation that implies the emotion distribution of the phoneme sequence, allowing for accurate and natural speech synthesis. Inspired by  Deep Factorization \cite{lantian_li_2018_deep}, we disentangle emotion manifolds from reference emotional speech through Vector-Quantized VAE \cite{oord_2017_vqvae} reconstruction, using paired emotion-free speech as the "linguistic part" and speaker identity as the "speaker part". Unlike prosody- or style-related representations, our visualization experiments show that the emotion manifold has a linear relationship with emotional intensity. 
Furthermore, to obtain a more accurate explicit emotion distribution on the phoneme sequence, We propose the SWER (i.e. \textbf{\underline{S}}liding \textbf{\underline{W}}indow \textbf{\underline{E}}motion \textbf{\underline{R}}ecognizer). The SWER is a convolutional neural network equipped with the sliding window mechanism, which enhances the temporal context of the input when recognizing emotions at the phoneme level, thus addressing the issue of emotions not being present within individual phonemes. By cascading the proposed SWER with the Emotion Manifold learning module, we bridge the semantic gap between the synthesis model and explicit emotion labels. Our CASEIN framework demonstrates significant advantages over reproducible SoTA methods \cite{kun_zhou_2022_mixedemo,yi_lei_2021_finegrained} in extensive experiments on fine-grained emotion intensity regulation and mixed emotions.


\begin{figure*}
  \begin{center}
  \includegraphics[width=0.98\textwidth]{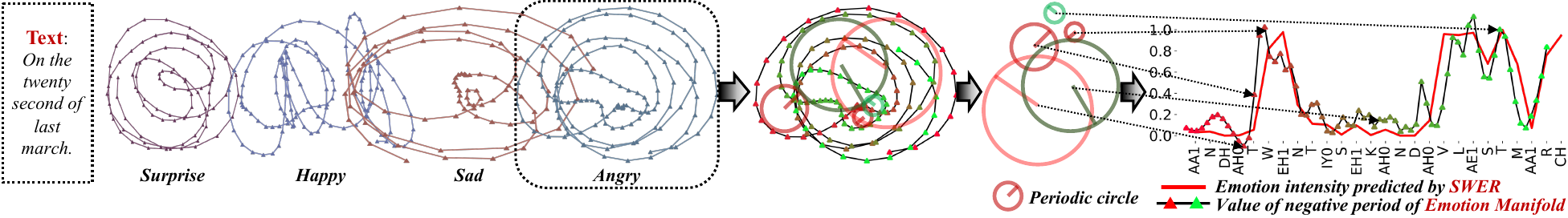}
  \vspace{-8pt}
  \caption{Visualization of the Emotion Manifold. The left side shows the manifold compressed by PCA, and the right side visualizes the period of each point on the manifold. The circumference of a periodic circle represents the period.}
  \vspace{-25pt}
  \label{fig:manishow}
  \end{center}
\end{figure*}

\section{Cascaded control learning}
\label{sec:cascaded_control_learning}


This section is dedicated to discussing the learning process of the explicit control module Emotion Manifold (Figure \ref{fig:framework}(a)) and the implicit control module SWER (Figure \ref{fig:framework}(c)).
The following section will detail the integration of these two modules in training the speech synthesis model CASEIN.

\subsection{Implicit Emotion Manifold learning}
\label{sec:implicit_emotion_manifold_learning}

Emotion manifold learning aims to extract emotion-related information from emotional speech mel-spectrograms $M_e$ at a fine-grained level. Disentangling the linguistic and speaker parts is necessary to obtain the emotional part, following the Deep Factorization \cite{lantian_li_2018_deep}. We use speaker identity embedding for the speaker part and paired\footnote{The paired refers to two speeches with the same text and speaker.} mel-spectrogram of emotion-free speech $M_n$ instead of text for the linguistic part. This improves linguistic part disentanglement accuracy in the mel-spectrogram space. VQVAE \cite{oord_2017_vqvae} is employed to disentangle each part, ensuring that the learned emotion part accurately represents the original mel-spectrogram.


First, we briefly introduce the Variational Autoencoder with Vector Quantization (VQVAE) framework. Given a feature to be reconstructed $X$, the continuous latent variables $z_c$ are obtained by encoding $X$ using an encoder $f_{enc}$. A codebook $e$ is constructed by setting $b$ clusters denoted by $e^1$, $e^2$, ..., $e^b$, and clustering $z_c$ on $e$ yields the discrete latent variables $z_d$. The objective is to reconstruct $z_d$ back to $X$ using a decoder $f_{dec}$. To perform disentanglement using VQVAE, we additionally include a condition $C$ to be reconstructed with $z_d$, enabling $z_d$ to learn a distinct part from $C$. 

In our framework, $X$ corresponds to the mel-spectrogram $M_e$. We construct the manifold predictor, denoted as \textbf{Pred M}, which consists of two convolutional layers serving as the $f_{enc}$\footnote{To align the length of the phoneme sequence with that of the mel-spectrogram, duration pooling is applied after the last layer.}. By inputting $M_e$ to Pred M and clustering, we obtain the latent variables $z_d$. We adopt the decoder from Fastspeech2 as $f_{dec}$ to accelerate the convergence. The condition $C$ includes the linguistic and speaker parts, as our objective is to disentangle emotion by removing these parts. The speaker part is obtained by embedding the speaker identity of $M_e$, while the linguistic part is extracted from the pronunciation of the emotion-free speech mel-spectrogram $M_n$. The extractor comprises two convolutional layers, followed by average pooling to retain only crucial information, such as pronunciation. Our customized VAE allows us to accurately learn the fine-grained implicit representation $z_d$, which is highly associated with emotion. As the learned $z_d$ exhibits continuously varying emotional periods, as shown in the visualization section, we refer to it as \textbf{Emotion Manifold}.

\subsection{Explicit SWER learning}
\label{sec:explicit_swer_learning}

The Sliding Window Emotion Recognizer aims to extract the emotion probability of each phoneme explicitly, which requires aligning the corresponding portion of each phoneme in $M_e$. Following the approach in Fastspeech2 \cite{yi_ren_2021_fastspeech2}, we use the Montreal forced alignment (MFA) \cite{michael_mcauliffe_2017_mfa} tool to obtain the boundaries for each phoneme. We then use these boundaries to slice $M_e$ into phoneme-level $M_p$ representations denoted by $M_p^1$, $M_p^2$, ..., $M_p^t$. However, as shown in our previous analysis, the length of the spectrum contained in $M_p^i$ is too short, resulting in low emotion classification accuracy. To address this issue, we propose using the Sliding Window technique to increase the spectrum length of each phoneme while simultaneously smoothing the distribution of the predictions. Specifically, we obtain $M_s$ by mapping sliding windows of radius $w$ to each phoneme in $M_p^i$. The $M_s^i$, which represents the i-th element of $M_s$, is the spectrum that contains the range from $M_p^{i-w}$ to $M_p^{i+w}$ centered at $M_p^i$.
During training, we treat each element in $M_s$ as an individual sample to train the emotion classification model. The class label $p\in\mathbb{R}^{n}$ represents the emotion category of $M_e$, where $n$ is the number of emotion categories. We introduce an emotion distribution predictor, referred to as \textbf{Pred D}, to predict the emotion probability $\tilde{p}$ of $M_s^i$. The predictor consists of two convolutional layers and one average pooling layer. To calculate the loss between $p$ and $\tilde{p}$, we use element-wise binary cross-entropy.
With the learned Pred D, we are able to classify each sliding window of the phoneme sequence and generate a fine-grained distribution of emotions denoted by $D\in\mathbb{R}^{t \times n}$.

\begin{table}
  \footnotesize
  \centering
  \tabcolsep=0.4cm
  \renewcommand\arraystretch{0.8}
  \caption{MCD and MOS score of the emotion restoration task. MCD is the lower the better and MOS is the higher the better.}
  \vspace{-7pt}
  \label{table:mos}
  \begin{tabular}{l|cc}
    \toprule
      model & MOS $\uparrow$ & MCD $\downarrow$ \\
    \midrule
      GT. & $4.82 \pm 0.05 $ & / \\
      MixedEmotion \cite{kun_zhou_2022_mixedemo} & $3.50 \pm 0.21$ & $5.66$  \\
      FET \cite{yi_lei_2021_finegrained} & $3.71 \pm 0.13$  & $5.15$ \\
    \midrule
      CASEIN (proposed) & $\mathbf{4.02 \pm 0.15}$ & $\mathbf{4.52}$ \\
    \bottomrule
  \end{tabular}
  \vspace{-7pt}
\end{table}

\section{Framework of CASEIN}
\label{sec:framework_of_casein}


This section describes the process of cascading the pre-trained explicit and implicit control modules to control speech synthesis. Our speech synthesis pipeline, as shown in Figure \ref{fig:framework}(b), is nearly identical to Fastspeech2, with the addition of an emotion adapter that takes as input the Emotion Manifold containing rich and accurate emotion information.
Assuming that the input phoneme sequence of the synthesis model is $T$ and the synthesis target is mel-spectrogram $M_e$. 
As the emotion manifold $z_d$ cannot be directly extracted during inference, we use $\tilde{z_d}$ prediction as a proxy task to train the implicit control module. To obtain $z_d$, we input $M_e$ to the pre-trained \textit{Pred M} $f_{predM}$ (described in \S\ref{sec:implicit_emotion_manifold_learning}), and cluster the output to obtain $z_d$. On the other hand, $\tilde{z_d}$ is generated by the emotion distribution $D$, which can be predicted by the explicit control module during training or manually input by the user during inference. The emotion distribution $D$ is transformed to the emotion manifold $\tilde{z_d}$ using a two-layer convolutional generator $f_{gen}$. The training objective of the implicit control can be formulated as follows.
\begin{equation}\begin{split}
  z^{i}_d=e^{k}, \mathrm{\ } k=\mathrm{argmin}_j\Vert f_{predM}(M_e)^{i}-e^j \Vert_2 \\
  \tilde{z}^{i}_d=e^{k}, \mathrm{\ } k=\mathrm{argmin}_j\Vert f_{gen}(D)^{i}-e^j \Vert_2 \\
  \mathcal{L}_{imp} (\tilde{z}_d, z_d) = \frac{1}{t} \sum_{i=1}^{t} \Vert \tilde{z}^{i}_d - z^{i}_d \Vert_2
\end{split}\end{equation}
where $t$ denotes the length of the $T$ and $e$ denotes the codebook.

Predicting the emotion distribution $D$ is a crucial step in training, as it determines the accuracy of the predicted emotion manifold and the ability to control synthesis with user input during inference. Following the SWER learning, we align $M_e$ to the phoneme boundaries of $T$, extract sliding windows corresponding to each phoneme, and obtain $M_s$. Applying the pre-trained \textit{Pred D} $f_{predD}$ (described in \S\ref{sec:explicit_swer_learning}) to each window in $M_s$ yields $D$ (denoted by $D^1$, $D^2$, ..., $D^t$) as $D^i = f_{predD}(M_s^i)$.

The predicted emotion manifold $\tilde{z_d}$ is fed into the two-layer convolutional emotion adapter $f_{ada}$ and then converged with the synthesis model $f_{syn}$ (i.e., Fastspeech2) to generate the mel-spectrogram $\tilde{M_e}$. The synthetic loss
is the same as the original Fastspeech2 and can be represented as follows.
\begin{equation}\begin{split}
  \tilde{M_e} = f_{syn}(T, f_{ada}(\tilde{z_d})) \\
  \mathcal{L}_{syn} (\tilde{M}_e, M_e) = \frac{1}{v} \sum_{i=1}^{v} \Vert \tilde{M}^{i}_e - M^{i}_e \Vert_2
\end{split}\end{equation}
where $v$ denotes the length of the mel-spectrogram $M_e$. Eventually the loss of CASEIN training can be expressed as $\mathcal{L}_{casein} = \mathcal{L}_{syn} + \lambda \mathcal{L}_{imp}$, where $\lambda$ is empirically set to 0.1.


\begin{figure}
  \begin{center}
  \includegraphics[width=0.4\textwidth]{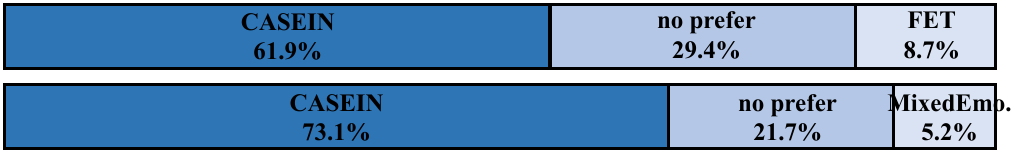}
  \vspace{-8pt}
  \caption{ABX preference test of the emotion restoration task.}
  \vspace{-30pt}
  \label{fig:abx}
  \end{center}
\end{figure}

\section{Experiments}
\label{sec:experiments}
\subsection{Evaluation setup}


Consistent with prior research \cite{kun_zhou_2022_mixedemo,yiwei_guo_2022_emodiff}, we trained our CASEIN framework using the English segment of the Emotional Speech Dataset (ESD) \cite{kun_zhou_2022_esd}. This dataset consists of 300 training texts, 20 validation texts, and 30 testing texts, each spoken by 10 English speakers expressing 5 distinct emotions: Neutral, Happy, Sad, Angry, and Surprised. For our experiment, we focused on a subset of the ESD dataset called '0015', which consists of a total of 1750 utterances from a female speaker. This subset is one of the most frequently used subsets in the ESD dataset. We extracted mel-spectrograms using parameters of 80 channels, 256 hop length and 16000Hz sampling rate.

Our training pipeline consists of three phases: emotion manifold learning, SWER learning, and CASEIN training. In the emotion manifold learning phase, we treat the Neutral as emotion-free speech and the other four emotions as emotional speech. We set the number of clusters to 256, with each cluster having a dimension of 512. The second convolutional layer of the \textit{Pred M} module maintains the same hidden size as the dimension of the cluster, while the other convolutional layers have hidden sizes, kernel sizes, and dropout ratios of 256, 9, and 0.2, respectively. In the SWER learning phase, we specify the sliding window size of 5. In the CASEIN training phase, we employ Fastspeech2
as our base model, with weights pre-trained on LibriTTS \cite{heiga_zen_2019_libritts}. HiFiGAN \cite{jungil_kong_2020_hifigan} is used as the vocoder for audio synthesis. We train the model using the Adam optimizer with a learning rate of 5e-4 and betas of (0.9,0.98). A linear learning rate decay scheduler is used, dropping from the maximum to 0 over 100 epochs. The entire training process takes 5 hours on a single Tesla V100 GPU.

We used objective and subjective metrics in our experiments, and 21 native speakers participated in the subjective evaluation. For objective metrics, we used Mean Cepstral Distortion (\textbf{MCD}) to estimate speech quality by calculating the MFCC of the synthesized speech and the ground truth speech. For subjective metrics, we used Mean Opinion Score (\textbf{MOS}), a 5-point scale widely used in TTS to assess speech quality. In addition, we conducted an \textbf{ABX} preference test where judges chose which speech sample was more similar to the reference in emotion expression, with three choices: (1) 1st is better; (2) 2nd is better; (3) no preference. To measure regulation ability in more detail, we proposed a Reference-Free ABX (\textbf{RF-ABX}) preference test, since the ABX metrics require reference audio, which may not always be available. In this test, judges rated which speech sample was more consistent with the given description of emotional expression.


We compared our method with the reproducible (SoTA) baseline methods. For the fine-grained emotional intensity regulation task, we used the \textbf{FET} method proposed by Lei et al. \cite{yi_lei_2021_finegrained}. For the emotions mixing task, we used the \textbf{MixedEmotion} method proposed by Zhou et al. \cite{kun_zhou_2022_mixedemo}. While some studies \cite{yi_lei_2022_msemotts,yiwei_guo_2022_emodiff} were proposed after the baseline methods, we did not use them as baselines since they are not open source and their contributions do not overlap with ours.

\begin{table}
  \footnotesize
  \centering
  \tabcolsep=0.2cm
  \renewcommand\arraystretch{0.8}
  \caption{The intensity curve from 0\% to 100\% means that the intensity of the first phoneme is 0\%, the last is 100\%, and each intermediate phoneme is obtained by linear interpolation.}
  \vspace{-10pt}
  \label{table:finegrained}
  \begin{tabular}{l|c|ccc}
    \toprule
      Emotion & Intensity Curve & CASEIN & no prefer & FET  \\
    \midrule
      \multirow{2}{*}{Happy} & 0\% - 100\% & \textbf{35\%} & 41\% & 24\% \\
       & 100\% - 0\% & 23\% & 48\% & \textbf{29\%} \\
    \midrule
      \multirow{2}{*}{Sad} & 0\% - 100\% & \textbf{51\%} & 31\% & 18\% \\
       & 100\% - 0\% & \textbf{48\%} & 28\% & 24\% \\
    \midrule
      \multirow{2}{*}{Angry} & 0\% - 100\% & \textbf{69\%} & 21\% & 10\% \\
       & 100\% - 0\% & \textbf{61\%} & 23\% & 16\% \\
    \midrule
      \multirow{2}{*}{Surprise} & 0\% - 100\% & \textbf{70\%} & 21\% & 9\% \\
       & 100\% - 0\% & \textbf{59\%} & 29\% & 12\% \\
    \bottomrule
  \end{tabular}
  \vspace{-5pt}
\end{table}

\begin{figure}
  \begin{center}
  \includegraphics[width=0.47\textwidth]{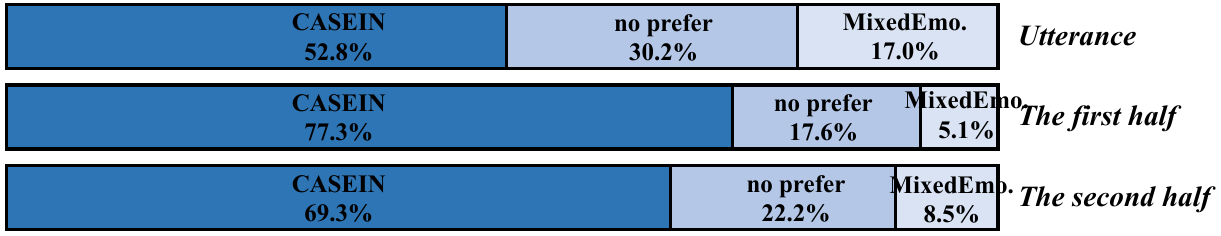}
  \vspace{-18pt}
  \caption{RF-ABX preference test of the emotion mixing task.}
  \vspace{-27pt}
  \label{fig:abx_mixed}
  \end{center}
\end{figure}



\subsection{Evaluation}


As shown in Figure \ref{fig:manishow}, we investigated the nature of the learned emotion manifold and found that it captures unique characteristics for each emotion. By extracting emotion manifolds from a pair of emotional speeches and projecting them onto a 2-dimensional space using PCA, we observed that the manifold for sadness is smoother than the manifold for happiness, demonstrating the manifold's ability to model the essence of emotions. We also observed a strong correlation between the fine-grained emotional intensity and the periodicity of the manifold. Using the shift of the tangent angle to approximate the period, we found that the negation of the computed period closely overlapped with the emotional intensity curve, with the former preserving more nuanced information. This discovery provides insight into the underlying mechanism of the emotion manifold's efficacy.


Our CASEIN model can regulate fine-grained emotion intensity while synthesizing more natural speech, as stated in section \S\ref{sec:introduction}. To validate this, we designed a speech restoration task where only the text and explicit emotion probabilities extracted from the target speech are input to synthesize the speech. We compared the performance of our model with SoTA models using objective and subjective metrics, as shown in Table \ref{table:mos} and Figure \ref{fig:abx}. Our model significantly outperformed the SoTA models in all three metrics, indicating that CASEIN can regulate emotions more accurately and synthesize higher-quality speech. 

\begin{table}
  \footnotesize
  \centering
  \tabcolsep=0.27cm
  \renewcommand\arraystretch{0.8}
  \caption{To convert A to B, the attenuation curve of A is combined with the strengthening curve of B.}
  \vspace{-10pt}
  \label{table:finegrained_mixed}
  \begin{tabular}{l|c|ccc}
    \toprule
      Convert A & To B & CASEIN & no prefer & FET  \\
    \midrule
      \multirow{3}{*}{Happy} & Sad & \textbf{57\%} & 25\% & 18\% \\
       & Angry & \textbf{78\%} & 6\% & 16\% \\
       & Surprise & \textbf{83\%} & 7\% & 10\% \\
    \midrule
      \multirow{3}{*}{Sad} & Happy & \textbf{64\%} & 17\% & 19\% \\
       & Angry & \textbf{75\%} & 13\% & 12\% \\
       & Surprise & \textbf{77\%} & 9\% & 14\% \\
    \midrule
      \multirow{3}{*}{Angry} & Happy & \textbf{85\%} & 6\% & 9\% \\
       & Sad & \textbf{84\%} & 8\% & 8\% \\
       & Surprise & \textbf{88\%} & 4\% & 8\% \\
    \midrule
      \multirow{3}{*}{Surprise} & Happy & \textbf{60\%} & 22\% & 18\% \\
       & Sad & \textbf{71\%} & 15\% & 14\% \\
       & Angry & \textbf{78\%} & 16\% & 6\% \\
    \bottomrule
  \end{tabular}
  \vspace{-18pt}
\end{table}

\noindent\textit{\textbf{Fine-grained intensity regulation:}} We tested CASEIN's ability to regulate emotions with single primary emotions in a fine-grained intensity task, using RF-ABX to measure emotional strengthening and attenuation curves. Four emotions were assessed, and judges chose the voice that best represented the gradual strengthening or attenuation of the emotion. Results (see Table \ref{table:finegrained}) showed that CASEIN performed better for intense emotions like Angry and Surprise, with a smaller gap for moderate emotions. This is because intense emotions provide more information, but accurately modeling the explicit emotion distribution at the phoneme level is difficult. CASEIN was found to be more effective in emotion strengthening than attenuation, likely due to more redundant data for emotion strengthening resulting in better emotion manifold modeling. This can be a limitation of CASEIN.






\noindent\textit{\textbf{Emotion mixing:}} Here we demonstrate the superiority of CASEIN in fine-grained emotion mixing. Our baseline models are MixedEmotion for coarse-grained mixing and FET for fine-grained mixing. We use equal emotion probabilities for each phoneme when comparing with the coarse-grained approach to ensure fairness. The experiments are conducted on the test set of the ESD dataset, and we regulate the intensity of each emotion to synthesize new emotions\footnote{We synthesized three new emotions, including Proud = 90\%Happy + 45\%Surprise; Disappointed = 70\%Sad + 64\%Angry; Devastated = 10\%Surprise + 93\%Sad.} (i.e. Proud, Disappointed, Devastated) based on psychological theories \cite{georgios_n_yannakakis_2017_theordinal}. To verify emotional consistency, we also conduct experiments on the first and second halves of the sentence. Using RF-ABX metrics, we find that our CASEIN can better express the new emotions, especially in terms of back-and-forth consistency, as shown in Figure \ref{fig:abx_mixed}. This suggests that coarse-grained methods tend to combine multiple emotions, leading to a lack of truly mixed emotions as the training objective is to express average sentiment.


\noindent\textit{\textbf{Fine-grained emotion conversion:}} 
We conducted emotion conversion as the task for comparison, where we combined the intensity curves of two emotions, such as converting emotion A to emotion B by regulating the intensity curve of A to 100\%-0\% and that of B to 0\%-100\%. The results presented in Table \ref{table:finegrained_mixed} demonstrate that our advantage in mixed emotion control is more pronounced compared to primary emotion control. This suggests that explicitly modeling mixed emotions at a fine-grained level is challenging and underscores the significant contribution of our cascaded control approach.

\section{Conclusions}
\label{sec:conclusions}


Our proposed framework, CASEIN, integrates explicit and implicit emotion control in a cascaded manner using two modules: the Emotion Manifold for implicit emotion modeling and SWER for explicit fine-grained emotion distribution extraction. By cascading these modules, we address the inaccuracy issue of existing explicit control methods in fine-grained emotion modeling, particularly for mixed emotions. We demonstrate the effectiveness of CASEIN in fine-grained emotion regulation and mixing tasks, where it can synthesize high-quality speech and enhance fine-grained emotion control capability.

\bibliographystyle{IEEEtran}
\bibliography{mybib}

\end{document}